\documentclass{article}

\usepackage{epsfig}  

\usepackage{url}  
\usepackage{latexsym,amssymb}  
\usepackage{boxedminipage}  
\usepackage{xcolor}

\usepackage[linesnumbered,longend]{algorithm2e}
\usepackage[round]{natbib}

\usepackage[left]{lineno}

\usepackage{etoolbox}

\begin{document}  
  \linenumbers
  

\newcommand{\mC}{\mathcal{C}}  
\newcommand{\mF}{\mathcal{F}}  
\newcommand{\sign}{\mbox{\it sign}}
\newcommand{\PK}{\mathcal{P}}
\newcommand{\n}{[n]}
\newcommand{\Id}{\mbox{\it Id}}
\newcommand{\found}{\it found}
\newcommand{\larg}{\it large}
\newtheorem{remark}{Remark} 

\newtheorem{definition}{Definition}  
\newtheorem{proposition}{Proposition}  
\newtheorem{prop}{Property}  
\newtheorem{lemma}{Lemma}  
\newtheorem{cor}{Corollary}  
\newtheorem{corollary}{Corollary}  
\newtheorem{example}{Example}  
\newtheorem{invariant}{Invariant}  
\newtheorem{property}{Property}  
 
\newtheorem{theo}{Theorem}  
\newtheorem{theorem}{Theorem}  

\newenvironment{proof2}{\noindent {\it Proof.}}{$\Box$\vskip1ex}  
\newenvironment{preuve}{\noindent {\it Proof.}}{$\Box$\vskip1ex}  
\newenvironment{preuveA1}{\noindent {\it Proof of conservation of   
Property $A_1$.}}{$\Box$\vskip1ex}  
\newenvironment{preuveA2}{\noindent {\it Proof of conservation of   
Property $A_2$.}}{$\Box$\vskip1ex}  
\newenvironment{preuveB}{\noindent {\it Proof of the invariants.}}  
{$\Box$\vskip1ex}  
\newenvironment{OJO}{\noindent {\bf OJO:}}{$\Box$\vskip1ex}  
  
  
\title{Easy identification of generalized common and conserved nested intervals}

\author{
Fabien de Montgolfier\thanks{\mbox{LIAFA, Univ. Paris Diderot - Paris 7, 75205 Paris Cedex 13, France.} {\tt \{fm,raffinot\}@liafa.univ-paris-diderot.fr}} \and Mathieu Raffinot$^*$ \and Irena Rusu\thanks{\mbox{LINA, UMR 6241, Universit\'e de Nantes, 44300, Nantes, France.} {\tt Irena.Rusu@univ-nantes.fr}}} 
\maketitle

\begin{abstract}

In this paper we explain how to easily compute gene clusters,
formalized by classical or generalized nested common or conserved
intervals, between a set of $K$ genomes represented as $K$
permutations. A $b$-nested common (resp. conserved) interval $I$ of
size $|I|$ is either an interval of size 1 or a common
(resp. conserved) interval that contains another $b$-nested common
(resp. conserved) interval of size at least $|I|-b$. When $b=1$, this
corresponds to the classical notion of nested interval. We exhibit two
simple algorithms to output all $b$-nested common or conserved
intervals between $K$ permutations in $O(Kn+\mbox{\em nocc})$ time,
where {\em nocc} is the total number of such intervals. We also
explain how to count all $b$-nested intervals in $O(Kn)$ time. New properties of the family of conserved intervals are proposed to do so.
\end{abstract}  
  


\section{Introduction}  
\label{sec:intro}


Comparative genomics is nowadays a classical field in computational
biology, and one of its typical problems is to cluster sets of
orthologous genes that have virtually the same function in several
genomes. A very strong paradigm is that groups of genes which remain
``close'' during evolution work together (see for instance \citet{galperin2000s}, \citet{lathe2000gene}, \citet{tamames2001evolution}).
Thus, a widely  used approach
to obtain interesting clusters is to try to cluster genes or other
biological units (for instance unique contigs of protein domains)
according to their common proximity on several genomes.  For this
goal, many different cluster models have been proposed, like common
intervals in \citet{UnoYagura}, conserved intervals in \citet{BergeronBCC04},
$\pi$-patterns in \citet{Parida06}, gene teams in \citet{BealBCR04}, domain
teams in \citet{Pasek2005}, approximate common intervals in \citet{AmirGS07}
and so on, considering different chromosome models (permutations,
signed permutations, sequences, graphs, etc.) or
different distance models (accepting gaps, distance modeled as
weighted graphs, etc.)

Among all those models, the first proposed, and still
one of the most used in practice, is the concept of \emph{common
  interval} on genomes represented by permutations. A set of genes
form a common interval of $K$ genomes if it appears as a segment on
each of the $K$ unsigned permutations that represent the genomes. The orders
inside the segments might be totally different.  The model of
conserved interval is close to the model of common interval but
considers signed permutations.

Recently, nested common intervals (easily extensible to nested
conserved intervals) were introduced in \citet{HD05} based on real data
observation by \citet{The10}. A common interval $I$ of size $|I|$ is nested 
if $|I|=1$ or if it contains at least one  nested common interval of size $|I|-1$.
\citet{HD05} pointed that
the nestedness assumption can strengthen the significance of detected
clusters since it reduces the probability of observing them randomly.
An $O(n^2)$ time algorithm to compute all nested common intervals
between two permutations has been presented in \citet{BlinFS10} while
between $K$ permutations a recent $O(Kn+\mbox{\em nocc})$ algorithm is
proposed in \citet{IR}, where \emph{nocc} is the number of solutions.

In this paper, we exhibit two simple algorithms to easily compute
nested common and conserved intervals of $K$ permutations from their
natural tree representations. Also, with the same simplicity, we
propose to deal with a generalization of nested common intervals,
called $b$-nested common intervals, and with its variant for conserved
intervals (which are a signed version of common intervals), called
$b$-nested conserved intervals.  These new classes allow - as $b$
grows - for a less constraint containment between the intervals in the
family. Indeed, a nested interval $I$ must contain a nested interval
of size $|I|-1$ or be a unit interval. A $b$-nested common
(resp. conserved) interval must contain a $b$-nested common
(resp. conserved) interval of size at least $|I|-b$ or be a unit
interval. Nested intervals are indead $1$-nested. 
From a biological point of view, this is equivalent to modeling
clusters with a larger variability in gene content and gene order,
thus allowing algorithms to deal with annotation errors. However, the
study and validation of this new interval model is deferred to
further applied studies. In this paper we focus on the algorithmic
aspects.

Given a set $\PK$ of $K$ permutations on $n$ elements representing
genomes with no duplicates, our simple algorithms 
for finding all $b$-nested common or conserved intervals of $\PK$ run in
$O(Kn+\mbox{nocc})$-time and need $O(n)$ additional space, where
\emph{nocc} is the number of solutions. In this way, our algorithm for 
common intervals performs as well as the algorithm in \citet{IR} for the case
of $K$ permutations, and proposes an efficient approach for the new classes of 
$b$-nested intervals.
 
Moreover, a slight modification of our approach allows us to count the number
of $b$-nested common or conserved intervals  of $K$ permutations in $O(Kn)$ time.
Efficiently counting the number of $b$-nested common intervals without enumerating
them is very usefull when one needs to compute similarity functions between genomes that
are expressed in terms of number of intervals.  See for instance \citet{FR}.

The paper is organized as follows. In Section \ref{sect:Generalities}
we present the main definitions for common and conserved intervals, and
precisely state the problem to solve. In Section \ref{sect:PQ} we
focus on $b$-nested common intervals, recalling the data structure called a 
$PQ$-tree, giving a characterization of $b$-nested common intervals and showing 
how $PQ$-trees can be used to find all $b$-nested common intervals. 
In Section \ref{sect:conservedtree} we adopt a similar approach for
conserved intervals, with the difference that another tree structure must
be  used in this case. In Section \ref{sect:conclusion} we eventually conclude.

\section{Generalities on common and conserved intervals}\label{sect:Generalities}
A {\em permutation} $P$ on $n$ elements is a complete linear order on the set
of integers $\{1, 2, \ldots, n\}$. We denote $Id_n$ the identity
permutation $(1, 2, \ldots, n)$. An {\em interval} of a permutation
$P = (p_1, p_2, \ldots, p_n)$ is a set of consecutive elements of the
permutation $P$. An interval of a permutation will be denoted 
by  giving its first and last positions, such as $[i,j]$. Such
an interval is also said {\it delimited} by $p_i$ (left) and $p_j$ (right).
An interval $[i,j] = \{i, i+1, \ldots, j\}$ of the identity permutation 
will be simply denoted by $(i .. j)$.

\begin{definition}[\citet{UnoYagura}]
  \label{def:common-intervals}
  Let $\mathcal{P} = \{P_1, P_2, \ldots, P_K\}$ be a set of $K$ permutations
  on $n$ elements.
  A {\em common interval} of  $\mathcal{P}$ 
  is a set of integers that 
  is an interval in each permutation of  $\mathcal{P}$.
\end{definition}

The set $\{1, 2, \ldots, n\}$ and all singletons (also called unit
intervals) are common intervals of any non-empty set $\PK$ of
permutations.  Moreover, one can always assume that one of the permutations,
say $P_1$, is the identity permutation $Id_n$. For this, it is sufficient
to renumber the elements of $P_1$ so as to obtain $Id_n$, and then 
to renumber all the other permutations accordingly. Then the common
intervals of $\PK$ are of the form $(i..j)$ with $1\leq i\leq j\leq n$.

Define now a {\it signed} permutation as a permutation $P$ whose elements have an associated sign among
 $+$ and $-$, making each element to be respectively {\em positive} or {\em negative}.  
Negative elements are denoted $-p_i$ while positive elements are simply denoted $p_i$, or $+p_i$ for emphasizing positivity. A permutation is then a signed permutation containing only  positive elements. 

\begin{definition}[\citet{BS06}]
Let $\PK =\{P_1, P_2,$ $\ldots,$ $P_K\}$ be 
a set of signed permutations over $\{1, 2, \ldots, n\}$, with first element $+1$ and last element $+n$, 
for each $k\in\{1, 2, \ldots, K\}$. Assume $P_1=\Id_n$.
A {\em conserved interval} of $\PK$ is either a unit interval or a common interval $(a..c)$ of $\PK$
(ignoring the signs) which is delimited, in each $P_k$, either by $a$ (left) and $c$ (right),
or by   $-c$ (left) and $-a$ (right).
\label{def:conserved}
\end{definition}

\begin{remark}
In the subsequent, we assume that $\PK =\{P_1, P_2,$ $\ldots,$ $P_K\}$ with $P_1=Id_n$. Moreover,
when we deal with conserved intervals, the permutations are assumed to satisfy the hypothesis in
Definition \ref{def:conserved}. 
\end{remark}

Now, we are ready to introduce the new classes of intervals.

\begin{definition}
  \label{def:bnested-intervals}
  Let $\PK = \{P_1, P_2, \ldots, P_K\}$ be a set of $K$ permutations
  on $n$ elements and let $b$ be a positive integer. 
  A common  (respectively conserved) interval of  $\PK$ 
  is {\em $b$-nested} if either $|I|=1$ or $I$ strictly contains a common (resp. conserved)  interval of size at 
  least $|I|-b$.
\end{definition}

We are interested in efficient algorithms for finding and counting all 
$b$-nested common  (resp. conserved)
intervals of $\PK = \{P_1, P_2, \ldots, P_K\}$, without redundancy. 
Obviously,
unit intervals are, by definition, $b$-nested common (resp. conserved) intervals.
As a consequence, from now on and without any subsequent specification, we 
focus on finding $b$-nested common (resp. conserved)  intervals of size at
least 2.
The following notions will be very useful in the subsequent.

\begin{definition}
 Let $\mathcal{P} = \{P_1, P_2, \ldots, P_K\}$ be a set of $K$ permutations
  on $n$ elements and let $b$ be a positive integer. 
  A common  (resp. conserved) interval of  $\PK$ 
  is {\em $b$-small} if its size does not exceed $b$. Otherwise, the interval is {\em $b$-large}.
\end{definition}

Notice that all $b$-small intervals are $b$-nested, by definition and since unit intervals are $b$-nested.

\section{On  $b$-nested common intervals}\label{sect:PQ}

This section is divided into three parts. The first one recalls a tree structure that we associate
to common intervals of permutations, the $PQ$-trees. The second one discusses the properties of
$b$-nested common intervals. Finally, we give the algorithms for efficiently computing and counting
the $b$-nested common intervals.
 
\subsection{$PQ$-trees and common intervals}

\begin{definition} Let $\mathcal{F}$ be a family of intervals from $Id_n$ containing the interval $(1..n)$. 
A {\em PQ-tree representing the family $\mF$} is a tree $T(\mF)$ satisfying:

\begin{itemize}
\item[1.] its nodes are in bijection with a subset $S(\mF)$ of intervals from $\mF$, the root corresponding to $(1..n)$
\item[2.] its arcs represent all the direct (not obtained by transitivity) inclusions between intervals in $S(\mF)$
\item[3.] each node is labeled $P$ ou $Q$, and an order is defined for the children of each $Q$-node
\item[4.] an interval $I$ of $Id_n$ belongs to $\mF$ iff either it corresponds to a node, or there
exists a unique $Q$-node $z$ such that $I$  is the union of intervals corresponding to
successive children of $z$, according to the order defined for $z$.
\end{itemize}
\label{def:PQ}
\end{definition}

Note that the size of the tree is in $O(|S(\mF)|)$, thus allowing to drastically reduce the memory
space needed to store all the intervals in $\mF$. When labels $P$ and $Q$ are forgotten, the tree $T(\mF)$ 
is called the {\em inclusion tree} of $S(\mF)$.

Given the $PQ$-tree representing a family $\mF$, we denote by $Int(x)$ the interval from $S(\mF)$ corresponding
to a node $x$. We also denote, for each interval $I$ from $\mF$, by $D(I)$ the {\it domain} of $I$ defined 
as follows. If $I\in S(\mF)$, then $D(I)$ is the set of its children. If $I\not\in S(\mF)$, then by condition
4. in Definition \ref{def:PQ}, let $x_l, x_{l+1}, \ldots, x_r$ be the children of the $Q$-node $z$ such
that $I=\cup_{i\in(l..r)}Int(x_i)$. Then $D(I)=\{x_l, x_{l+1}, \ldots, x_r\}$.

Fundamental results on $PQ$-trees involve {\it closed} families of intervals.

\begin{definition}
A {\em closed} family $\mathcal F$ of intervals of the permutation $Id_n$ is a family that contains all singletons as well  
as the interval $(1..n)$, and that in addition has the following property:  
if $(i..k)$ and $(j..l)$, with $i\le j \le k \le l$, belong to $\mathcal F$, then  
$(i..j-1)$, $(j..k)$, $(k+1..l)$  and $(i..l)$ belong to $\mathcal{F}$.
\end{definition}

The construction of a $PQ$-tree for a closed family of intervals relies on strong
intervals:

\begin{definition}
Let $\mF$ be a family of closed intervals from $Id_n$. An interval $(i..j)$ is said to \emph{overlap} another interval 
$(k..l)$ if they intersect without inclusion,  \emph{i.e.} $i<k\le j<l$ or $k<i\le l<j$. 
An interval $I$ of $\mathcal F$ is {\em strong} if it does not overlap any other  interval of $\mF$, 
and  is {\em weak} otherwise.
\label{def:strong}
\end{definition}

Notice that  $(1..n)$ and the unit intervals are always strong. Also,  the family 
of strong  intervals of $\mF$ is \emph{laminar} (that is, 
every two distinct intervals are either disjoint or included in each other) 
and, as $(1..n)$ belongs to the family, it is possible to define for them an inclusion tree.
Then it can be shown that:

\begin{theorem}[\citet{landau2005using}]
\label{th1}
Given a closed family ${\mF}$ of  intervals of $Id_n$, let $S(\mF)$ be the set of strong intervals from $\mF$ and let
$T(\mF)$ be the inclusion tree of $S(\mF)$. Then the $PQ$-tree obtained by the following rules 
represents the family $\mF$:
\begin{itemize}
\item[1.] label with $P$ each node $x$ of $T(\mF)$ such that $\cup_{z\in D'}Int(z)\not\in\mF$ for all 
$D'\subset D(Int(x))$  with $2\leq |D'|<|D|$.
\item[2.] label with $Q$ each node $y$ of $T(\mF)$ not labeled $P$, and define the order $y_1, y_2, \ldots, y_r$
of its children such that $max(Int(y_i))<min(Int(y_{i+1}))$ for all $i<r$.  
\end{itemize}

\end{theorem}

Common intervals of permutations (including those of size 1) are obviously a closed family of intervals from $Id_n$, thus 
Theorem \ref{th1} applies. 
Moreover, the $PQ$-tree for common intervals (hereafter simply denoted $T$) may be computed in linear time:

\begin{theorem}\citet{BCMR05}
The construction of the $PQ$-tree $T$ of common  intervals of a set $\PK$ of
$K$ permutations on $n$ elements may be done in $O(Kn)$ time.
\end{theorem}

\begin{figure}[t]
 \begin{center}
\begin{tabular}{p{4cm}lp{4cm}} 
\vspace {-.8 cm}\includegraphics[height=3cm]{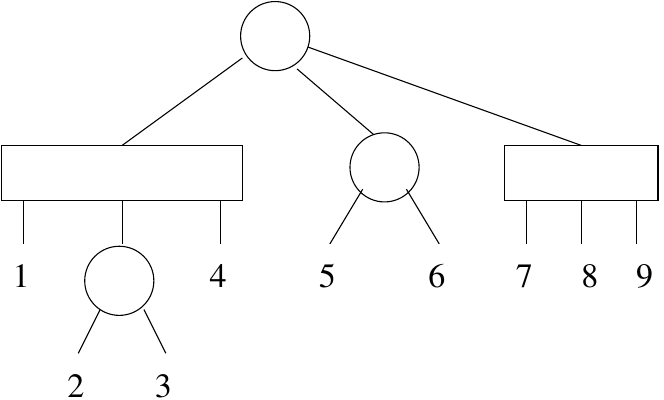}&
\phantom{xxxxxxxx} & 
\mbox{  Let $\mathcal{P}   = \{Id_9,P_2,P_3\}$,}\hfill \linebreak \centerline{$P_2=(4,2,3,1,7,8,9,6,5)$} \centerline{$P_3=(5,6,1,3,2,4,9,8,7)$}
 The $PQ$-tree for the set of common intervals of $\mathcal{P}$ is shown on the left.
\\
\end{tabular}
 \end{center}\vspace{-.3cm}
\caption{Example of PQ-tree}
 \label{figtotal2b}
\end{figure}

It is easy to note here that the leaves of $T$ are the singletons. Also,  the intervals
$Int(y_i)$, $i=1, 2, \ldots, r$, associated to the children of a $Q$-node are contiguous, 
i.e. $max(Int(y_i))+1=min(Int(y_{i+1}))$. This is due to condition 4 in the definition of a
$PQ$-tree and to the assumption, that the reader must keep in mind, that $P_1=Id_n$ (thus
all the common intervals are of the form $(i..j)$). An example is given in Fig. \ref{figtotal2b}.

\subsection{Properties of $b$-nested common intervals}\label{sect:main}

Let $\PK=\{P_1, P_2, \ldots, P_K\}$ be a set of permutations on $n$ elements such that $P_1=Id_n$, and
let $T$ be the $PQ$-tree representing the common intervals of $\PK$.
Say that a common interval $I$ is a {\em $P$-interval} if it is strong and there is a $P$-node $x$ with $Int(x)=I$. 
Otherwise, $I$ is a  {\em $Q$-interval}.

With the aim of identifying the particular structure of $b$-nested common  intervals
among all common  intervals, we first prove that:

\begin{lemma}\label{lemma:nonnested}
Let $I$ be a $b$-nested common  interval with $D(I)=\{x_1, x_2, \ldots, 
x_r\}$, $r\geq 1$. Then each of the intervals $Int(x_i)$, $1\leq i\leq r$,  is either a $b$-small 
or a $b$-nested common interval.
\end{lemma}
\begin{preuve}
Assume a contrario that some $Int(x_i)$ is of size $u\geq b+1$ and is not $b$-nested. 
Let $I'\subseteq I$ be a $b$-nested common  interval with the property 
that $x_i\in D(I')\subseteq D(I)$ and $D(I')$ is minimal with this property. Now,
since $I'$ is $b$-nested, we have that $I'$ strictly contains $Int(x_i)$ and thus
$|I'|>1$. Then $I'$ must contain some $b$-nested common  interval $J$ with
 $|I'|>|J|\geq|I'|-b$. Furthermore, $J$ and $Int(x_i)$ are disjoint since $Int(x_i)$ is strong
 and by the minimality of $I'$
we have that $J$ cannot contain $Int(x_i)$. But then $|I'|\geq |J|+|Int(x_i)|\geq |I'|-b + b+1=|I'|+1$,
a contradiction.
\end{preuve}

It is easy to see that:

\begin{remark}\label{rem:inclusion}
{\em Let $I, L, J$ be common  intervals such that $J\subseteq L\subseteq I$ and
$J$ is $b$-nested with $|J|\geq |I|-b$. Then  $L$ is $b$-nested, since
$|J| \geq |I|-b\geq |L|-b$.}
\end{remark}

Now, the characterization of $b$-nested intervals corresponding to a $P$-node is obtained as follows.

\begin{lemma}\label{lemma:round}
 Let $I$ be a $P$-interval with $D(I)=\{x_1, x_2, \ldots, x_r\}$. Then $I$ is a $b$-nested common 
 interval if and only if there is some $i, 1\leq i\leq r$, such that $Int(x_i)$ is
a $b$-nested common  interval of size at least $|I|-b$.
\end{lemma}

\begin{preuve}
Since $I$ is a $P$-interval, its maximal common  subintervals are $Int(x_i)$,
$1\leq i\leq r$. The ''$\Leftarrow$'' part  follows directly from the definition. For the
''$\Rightarrow$'' part, assume by contradiction that the affirmation does not hold. Then none of 
the intervals $Int(x_i)$, $1\leq i\leq r$, is $b$-nested of size at least $|I|-b$, but 
since $T$ is $b$-nested we deduce that some interval $Int(x_i)$ exists
containing a $b$-nested common  interval $J$ of size at least $|I|-b$. But
this is impossible according to Remark \ref{rem:inclusion}.
\end{preuve}

\begin{figure}[t]
\vspace{-0.3cm}
  \centering
\includegraphics[width=5cm]{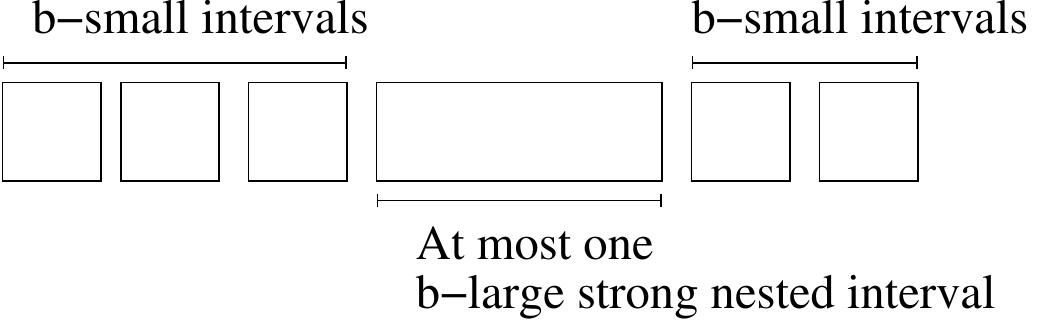} 
\caption{Structure of a $b$-nested $Q$-interval. }
 \label{bstruct}
\vspace{-0.3cm}
\end{figure}

The structure of $b$-nested common intervals given by consecutive children of a $Q$-node is more
complex. In the next lemmas we show that at most one of the intervals $Int(x_i)$ composing
such an interval may be $b$-large (see also Fig. \ref{bstruct}).
 
\begin{lemma}\label{lemma:firstlast}
 Let $I$ be a $Q$-interval with $D(I)=\{x_1, x_2, \ldots, x_r\}$. Then $I$ is a $b$-nested common 
 interval if and only if 
 $Int(x_1)$ is  $b$-small and $I-Int(x_1)$ is a $b$-nested common  interval, or $Int(x_r)$ is
 $b$-small and $I-Int(x_r)$ is a $b$-nested common  interval.
\end{lemma}

\begin{preuve}
Recall that for a $Q$-interval $I$, the order $x_1, x_2, \ldots, x_r$ implies that the equation
$max(Int(x_i))+1=min(Int(x_{i+1}))$ holds
for all $i$ with $1\leq i<r$. 

$\Rightarrow$: Since $I$ is $b$-nested, it contains some $b$-nested interval $J$ such that $|I|>|J|\geq |I |-b$.
Choose $J$ as large as possible. Now, $J$ cannot be strictly included in some non $b$-nested $Int(x_i)$ by Remark \ref{rem:inclusion}, thus
$D(J)=\{x_p, x_{p+1}, \ldots, x_s\}$ with $p\geq 1, s\leq r$, and $p\neq 1$ or $s\neq r$.
Assume w.l.o.g. that $p>1$. Then $Int(x_1)$ is $b$-small (since $|J|\geq |I |-b$)
and $I-Int(x_1)$ is $b$-nested by Remark \ref{rem:inclusion} since it contains $J$ or is
equal to $J$.

$\Leftarrow$: Let $j=1$ or $j=r$ according to which proposition holds. We have that $|I-Int(x_j)|=|I|-|Int(x_j)|\geq |I|-b$ since $|Int(x_j)|\leq b$. 
Then $I$ is $b$-nested.
\end{preuve}

\begin{lemma}\label{lemma:not2large}
 Let $I$ be a $Q$-interval with $D(I)=\{x_1, x_2, \ldots, x_r\}$ which is a $b$-nested common 
interval. Then at most one of the intervals $Int(x_i)$, $1\leq i\leq r$, is $b$-large, and in
this case this interval is a $b$-nested common  interval. 
\end{lemma}

\begin{preuve}
By contradiction, assume there exist $b$-nested common $Q$-intervals that contain at least two
$b$-large intervals of type $Int(x_i)$, and let $I$ be a smallest  such interval 
w.r.t. inclusion. Let $x_u$ (resp. $x_v$), with $1\leq u,v\leq r$, be such that
$Int(x_u)$ (resp. $Int(x_v)$) is $b$-large and $u$ (resp. $v$) is minimum (resp. maximum) with this
property. Then $u=1$ and $v=r$, otherwise by Lemma \ref{lemma:firstlast} the minimality of
$I$ is contradicted. But now Lemma \ref{lemma:firstlast} is contradicted, since $Int(x_1)$ and
$Int(x_r)$ are both $b$-large. 

Then, at most one of the intervals $Int(x_i)$, $1\leq i\leq r$, is $b$-large. To finish the proof, 
assume that $Int(x_i)$ (for some fixed $i$), is the unique $b$-large interval and 
apply Lemma \ref{lemma:nonnested} to $Int(x_i)$ to deduce that $Int(x_i)$ is $b$-nested.
\end{preuve}

We are able now to prove the theorem characterizing $b$-nested common  intervals.

\begin{theorem}\label{th:all}
Let $I$ be a common  interval of $\PK$.  $I$ is $b$-nested if and only if:
\begin{enumerate}
\item[(a)] either $I$ is a $P$-interval and there exists $x_h\in D(I)$ such that $Int(x_h)$ is a $b$-nested 
common  interval of size at least $|I|-b$.
\item[(b)] or $I$ is a $Q$-interval with the property that all intervals $Int(x_i)$ with $x_i\in D(I)$ are
 $b$-small, with one possible exception which is a $b$-large $b$-nested interval. 

\end{enumerate}
\end{theorem}

\begin{preuve}
Lemma \ref{lemma:round} proves the theorem in the case where $I$ is a $P$-interval. When
$I$ is a $Q$-interval, Lemma \ref{lemma:not2large} proves affirmation $(b)$.\end{preuve}

\subsection{Computing and counting all $b$-nested common intervals}\label{sect:algo}

\begin{figure}[t]
\vspace{-0.3cm}
  \centering
\includegraphics[width=9cm]{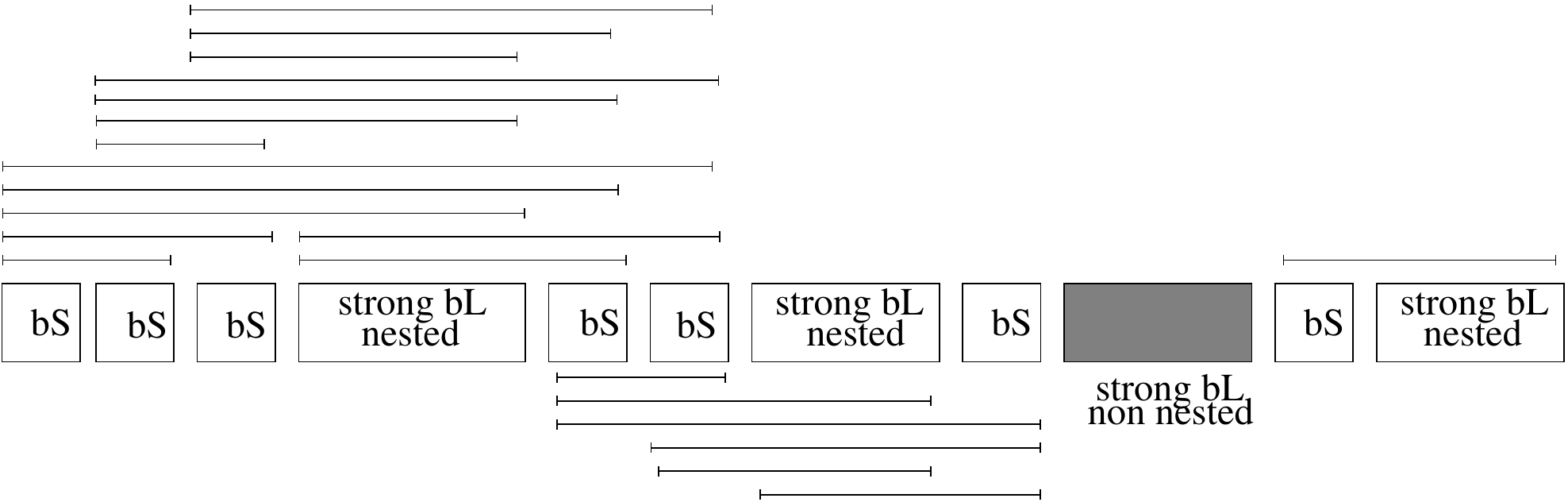} 
\caption{Computing all $b$-nested intervals of a $Q$-interval, where
  $bS$ (resp. $bL$) means $b$-small (resp. $b$-large). The algorithm
  considers all positions from left to right and expands the $b$-nested
  interval while it is possible.}
 \label{fig:algo}
\vspace{-0.3cm}
\end{figure}

Consider Algorithm \ref{algo:bnested1}, which computes all $b$-nested common intervals. 
For a node $x_c$, the  notations $min(c)$ and $max(c)$ respectively indicate the minimum and the maximum 
value in $Int(x_c)$.
Figure \ref{fig:algo} illustrates our algorithm.

\begin{algorithm}[h]
\label{algo:bnested1}
\dontprintsemicolon
\caption{The $b$-NestedCommonSearch algorithm}
\KwIn{The PQ-tree $T=(V,E)$ of $\PK$ for common  intervals, a positive integer $b$}
\KwOut{All $b$-nested common  intervals of $\PK$}
Perform a post-order traversal of $T$ \;
\For{each node $x$ of $T$ encountered during this traversal}{
  \eIf{$x$ is a leaf}{output $Int(x)$ as $b$-nested\;}{
     let $x_1, x_2, \ldots, x_p$ be the children of $x$ \;
     \eIf{$x$ is a $P$-node}{
      \If{$\exists$ $i$ such that $Int(x_i)$ is $b$-nested and $|Int(x_i)|\geq |Int(x)|-b$}{output $Int(x)$ as $b$-nested}
      }{
      \For{$c\leftarrow 1$ to $p$}{
           $\larg\leftarrow 0$ \hfill{//  number of $b$-large intervals already included}\;
           $d\leftarrow c$ \hfill{// considers all children starting with $x_c$}\; 
           \While{$d\leq p$ and ($|Int(x_d)|\leq b$ or $Int(x_d)$ is $b$-nested) and $\larg\leq 1$}{
             \lIf{$|Int(x_d)|>b$}{$\larg\leftarrow \larg+1 {\bf endif}$\;}
             \If{$c<d$ and $\larg\leq 1$}{
                output $(min(c)..max(d))$ as $b$-nested\;
                }
             $d\leftarrow d+1$\;
            }
         }
     }
}
}

\end{algorithm}

\begin{theorem}

Algorithm \ref{algo:bnested1} correctly computes all the $b$-nested common  intervals,
assuming the $PQ$-tree is already built, in $O(n+nocc)$ time, where $nocc$ is the
number of $b$-nested common intervals in $\PK$.
\end{theorem}

\begin{preuve} To show the algorithm correctness, note first that 
all the leaves are output in step 4, and they are $b$-nested common
intervals. Moreover, all $b$-nested common intervals corresponding to
$P$-nodes are correctly output in step 9 according to Theorem
\ref{th:all}$(a)$. Next, $Q$-intervals corresponding to a $Q$-node $x$
are generated in steps 12-22 by starting with each child $x_c$ of $x$,
and successively adding right children $x_d$ as long as condition
$(b)$ in Theorem \ref{th:all} is satisfied (step 15).

Let us analyze now the running time. The $PQ$-tree has size $O(n)$, and the traversal considers
every node $x$ exactly once. Working once on the children of each node
takes $O(n)$. The test in line 8 considers every child of a $P$-node
one more time, so that the $O(n)$ time is ensured when the $Q$-interval
generation is left apart. Now, during the generation of the
$Q$-intervals, a node $x_d$ that belongs to no $b$-nested common
interval is uselessly included in some interval candidate at most once
by left initial positions for the scan (beginning line 12), which is
in total bounded by $n$ since there exists a linear number of initial
positions in the $PQ$-tree. At each iteration of
the loop line 15, a unique distinct $b$-nested interval is output, or
$c=d$ (that happens once for each node since $d$ is incremented at
each iteration), or $large = 2$ (that also happens once for each node
since it ends the loop). The total number of iterations is thus
$O(n+\mbox{\em nocc})$, each iteration taking $O(1)$. The overall running 
time is thus in $O(n+\mbox{\em nocc}),$ where {\em nocc} is the
total number of $b$-nested common intervals.
\end{preuve}

The previous approach can be modified to count the $b$-nested common 
intervals instead of enumerating them, by simply analyzing more
precisely the structure of the $Q$-nodes. The goal is to count the $b$-nested
common intervals in a time proportional to the number of children instead of the
number of $b$-nested common intervals. To perform the count, we assume that the traversal is post-order, that it marks each
vertex as $b$-nested or not, and that it computes the cardinality of each $Int(x_i)$. 

{\bf $P$-nodes.} Obviously, a $P$-node (including the leaves) counts for 1 if the associated interval is $b$-nested, and for 0 otherwise.

{\bf $Q$-nodes.} To make the computation for a $Q$-node with children $\{x_1, x_2, \ldots,$ $x_p\}$, the algorithm looks for 
the $b$-large $b$-nested intervals $Int(x_d)$  and counts\\
1)~the $b$-nested common intervals containing each $Int(x_d)$,
and\\
2)~the $b$-nested common intervals generated by maximal sets of consecutive $b$-small common intervals 
$Int(x_i)$.  To this end, the vertices $x_i$ are considered from left to right in order to identify 
both the $b$-large $b$-nested common intervals $Int(x_d)$,
and the maximal sets of consecutive $b$-small common intervals 
$Int(x_i)$. Then: 

To solve 1), each encountered $b$-large $b$-nested interval $Int(x_d)$ has the following treatment. We 
count the number of consecutive $b$-small nodes $x_i$ on its right (resp. left), denoted by
$r(d)$ (resp. $l(d)$).  Then we compute the number of $b$-nested common intervals
which contain child $x_d$ as $$l(d)*(r(d)+1)+r(d).$$  

To solve 2),  for each maximal set of consecutive $b$-small common intervals 
$Int(x_i)$, assuming it has size $h$, we count $$h*(h-1)/2.$$ 

One may easily decide whether the interval corresponding
to the $Q$-node itself is $b$-nested or not, and compute its size. 

All these operations may be performed in $O(k)$ time, where $k$ is the number of sons of the $Q$-node.
\bigskip

{\bf Example.} On the
example in Fig. \ref{fig:algo}, we count (from left to right): (a)
for the first $b$-large strong child to the left: $3*(2+1)+2=11$ $b$-nested common
intervals; (b) for the $b$-large second strong child: $2*(1+1)+1=5$ $b$-nested common
intervals; (c) for the last $b$-large strong child: $1*(0+1)=1$ $b$-nested common
intervals. We sum up to obtain $17$ $b$-nested common intervals containing one $b$-large $b$-nested
interval $Int(x_i)$. Now we add the nested intervals generated between the $b$-large intervals: 
$3+1=4$. Altogether, the $Q$-node of Figure \ref{fig:algo} generates $21$ $b$-nested intervals.

\paragraph{Complexity} The time complexity of the counting procedure is obviously $O(n),$ the size of the underlying $PQ$-tree. The time needed to get the PQ-tree itself given $K$ permutations is however $O(Kn)$, as indicated before.



\section{On $b$-nested conserved intervals}
\label{sect:conservedtree}

As stated before, we assume the set $\PK$ of permutations has the properties required in
Definition \ref{def:conserved}. As conserved
intervals are common intervals, one may be tempted to follow the same approach using $PQ$-trees.
Unfortunately, the inclusion tree of strong conserved intervals does not define a $PQ$-tree representing
the family of conserved intervals, due to the fact that a  conserved interval  cannot be
written as a disjoint union of strong conserved intervals. The resulting 
{\em ordered tree} has been used in the literature \citet{BergeronBCC04}, but its underlying
properties have not been clearly stated. We do it here, before using these properties.

\subsection{Structure of conserved intervals}

We start by an easy property about intersection of conserved intervals:

\begin{lemma}\label{lemstru2}
Let $I=(u..v)$ and $J=(c..d)$ be two conserved intervals of $\PK$  with $u<c\le
v<d$. Then the intervals $(u..c)$, $(c..v)$, $(v..d)$ and
$(u..d)$ are conserved intervals.
\end{lemma}

\begin{preuve}
As an element $x$ from $I$ fulfills $u\le x\le v$ and an element $y$
from $J$ fulfills $c\le y \le d$ then an element $z$ from $I\setminus
J$ fulfills $u\le z < c$. These elements are exactly those between $u$
and $c$ and $(u..c)$ is thus a conserved interval. Similarly, 
$J\setminus I=(v..d)$.The elements from $I\cap J$ are all the elements 
not lower than $v$ and not larger than $c$,  so $(c..v)$ is a conserved interval.  
Finally, the elements from $I\cup J$ are 
all elements greater than $u$  and smaller than $d$, so $(u..d)$ is a conserved
interval. 
\end{preuve}

The notions of strong/weak intervals and of a frontier are essential in our study.

\begin{definition}
A conserved interval $I$ of $\PK$ is {\rm strong} if it has cardinality at least two, and does not overlap
other conserved intervals. Otherwise, it is {\em weak}.
\end{definition}

Notice that unit conserved intervals are not strong.

\begin{definition}
Let $I=(a..c)$ be a  conserved interval.  A set
$\{f_1, \ldots, f_k\}$ of elements satisfying $a=f_1<f_2<\ldots <f_k=c$ is a
\emph{set of frontiers} of $I$ if $(f_i..f_j)$ is a conserved
interval, for all $i,j$ with $1\le i < j \le k$. An element of $I$ is
a \emph{frontier} of $I$ if it occurs in at least one set of frontiers
of $I$.
\end{definition}

The two following properties are easy ones:

\begin{lemma}\label{lemstru1}
Let $I=(a..c)$ be a conserved interval and $F$ be a set of frontiers of $I$.
 The elements of $F$  are either all positive or all negative.
\end{lemma}
\begin{preuve}
 By the definition of a conserved interval, its two endpoints have the
 same sign. By the definition of a set on frontiers, any two frontiers 
 are the extremities of some conserved interval.
\end{preuve}

\begin{lemma}
\label{unionfrontier}
Let $I=(a..c)$ be a conserved interval, and  let $F=\{f_1, \ldots, f_k\}$ and
$F'=\{f'_1, \ldots, f'_l\}$ be two sets of frontiers of $I$. Then $F \cup F'$ is
also a set of frontiers of $I$.
\end{lemma}
\begin{preuve}
We show that any interval between two elements of $F\cup F'$ 
is conserved. Let
$f_i\in F$ and $f'_j\in F'$, and suppose that $f_i<f'_j$. If $f_i=a$, then we have
$(f_i..f'_j)=(f'_1..f'_j)$ and we are done.  If $f'_j=c$, then
$(f_i..f'_j)=(f_i..f_k)$ and we are also done. If $f_i\ne a$ and $f'_j\ne
c$, 
then Lemma \ref{lemstru2} allows to conclude.
The same proof holds if $f_i>f'_j$.
\end{preuve}

Let now $T$ be the inclusion tree of strong intervals from $\PK$, in which
every node $x$ corresponds to a strong interval denoted $Int(x)$, and 
node $x$ is the parent of node $y$ iff $Int(x)$ is the  smallest strong
conserved interval strictly containing $Int(y)$. Then $T$
contains two types of nodes: those corresponding to strong conserved 
intervals with no internal frontier, and those corresponding to strong 
conserved intervals with at least one internal frontier. We will show that 
weak conserved  intervals are the conserved strict subintervals of the latter ones, defined by two frontiers.
Overall, we have a structure working pretty much as a $PQ$-tree, but which 
cannot be mapped to a $PQ$-tree. This is proved in the next theorem.

Given a conserved interval, denote by $\mbox{Container}(I)$ 
the smallest strong conserved interval such that $I \subseteq J$.

\begin{theorem}
\label{theostruc2}
Each conserved interval $I$ of $\PK$ admits a unique maximal
(w.r.t. inclusion) set of frontiers denoted $F_I$.
Moreover, each conserved
interval $I$ of $\PK$ satisfies one of the following
properties:
\begin{itemize}
\item[1.] $I$ is strong
\item[2.] $I$ is weak and there exists a unique strong conserved interval $J$ of
  $\PK$, and two frontiers $f_i, f_j\in F_J$ with $f_i<f_j$, such that
  $I=(f_i..f_j)$. Moreover  $F_I = F_J\cap I$ and $J = \mbox{Container}(I).$
\end{itemize}
\end{theorem}

\begin{preuve}
Let us first prove the uniqueness of the maximal set of frontiers. By contradiction, assume
two distinct maximal sets of frontiers $F=\{f_1,\ldots,f_k\}$ and $F'=\{f'_1,\ldots,f'_l\}$ exist
for a conserved interval  $I$. Using Lemma \ref{unionfrontier}, we deduce that 
$F\cup F'$ is a larger set of frontiers of $I$, a contradiction.

Let us now prove that if $I$ is not strong, then  there exists a unique strong conserved interval
$J$, and two frontiers $f_i, f_j\in F_J$ with $f_i<f_j$, such that $I=(f_i..f_j)$.

{\em Existence of $J$.} Let $I_1=I=(a_0..a_1)$ be a weak conserved interval of $\PK$.
Then there is another interval $I_2$ overlapping it, either on its left
(i.e. $I_2=(a_2..x_2)$ with $a_2<a_0\le x_2 < a_1$)
or on its right (i.e. $I_2=(x_2..a_2)$ with $a_0<x_2\le a_1 <
a_2$). According to Lemma~\ref{lemstru2}, $J_2=I_1\cup I_2$ is a
conserved interval, and $F_2=\{a_0, a_1, a_2, x_2\}$ is a set of frontiers
for $J_2$ (not necessarily all distinct). If $J_2$ is not strong then it is overlapped by
another interval $I_3$.  We build an increasing sequence of intervals
$J_1=I_1,J_2,\ldots,J_k$, with $J_i$ overlapped by $I_{i+1}$, and
$J_{i+1} = J_i \cup I_{i+1}$, until we find a strong conserved
interval $J_k$ (the process ends since $(1..n)$ is strong). Each time
Lemma~\ref{lemstru2} ensures that $J_{i+1}$ is a conserved interval,
and $F_{i+1}=F_i\cup\{a_{i+1},x_{i+1}\}$ is a set of frontiers for $J_{i+1}$
(where $a_{i+1},x_{i+1}$ are the 
endpoins of  $I_{i+1}$). But then $F_{k}$ is a subset of the maximal set of
frontiers $F_{J_{k}}$ of $J_k$ and the two frontiers of  $J_k$ defining
$I_1$ are $a_0, a_1$, since $\{a_0,a_1\}\subseteq F_2\subseteq F_3\subseteq  \ldots\subseteq F_k\subseteq F_{J_k}$. 


{\em Uniqueness of $J$.} Assume a contrario that two strong intervals $J_1$ and $J_2$ exist with
$F_{J_1}=\{f_1, \ldots, f_p\}$, $F_{J_2}=\{f'_1, \ldots, f'_r\}$ and
such that $I=(f_i..f_j)=(f'_k..f'_l)$ with $1\leq i<j\leq p$ and
$1\leq k<l\leq r$.  Then, clearly, $f_i=f'_k$ and this is the left
endpoint of $I$, whereas $f_j=f'_l$ and this is the right endpoint of
$I$.  Since $J_1$ and $J_2$ are both strong, they cannot
overlap. Assume then w.l.o.g. that $J_1$ is strictly included in
$J_2$, and more precisely that $f_p\neq f'_r$ (then $f_p<f'_r$ due to
inclusion). Now, $(f'_k..f'_r)$ overlaps $J_1$ (which is forbidden) unless $i=1$, in which
case we have $f_1=f_i=f'_k$ and this is the left endpoint of $I$. Now,
since $I$ is strictly included in $J$, we deduce $j<p$ and
$(f'_l..f'_r)$ overlaps $J_1$. This contradiction proves the
uniqueness of the strong interval $J$ satisfying condition~2.

Let us prove that $F_I = F_J\cap I$. Suppose by contradiction that $F_I$
contains a frontier $f$ not in $F_J$. Recall that $I=(a_0..a_1)$ and 
$a_0$, $a_1$ belong to $F_J$. Now, for each $f_l\in F_J$ with $f_l < f$, we have that
$(f_l..f)$ is a conserved interval either by Lemma~\ref{lemstru2} applied
to $(a_0..f)$ and $(f_1..f_l)$ (when $f_l\geq a_0$) or since it is the union
of the two conserved intervals $(f_l..a_0)$ and $(a..f)$ (when $f_l<a_0$). 
Symmetrically, $(f..f_l)$ is also a conserved interval when $f_l\in F_J$, $f_l > f$. 
But then $F_J \cup \{f\}$ is a set of frontiers of $J$ larger than $F_J$, 
which contradicts the maximality of $F_J.$

Eventually, let us prove that $J = \mbox{Container}(I).$ Assume that
$F=\{f_1, f_2, \ldots,$ $f_k\}$, that $I=(f_i..f_j)$ and that, by contradiction, there 
is a smallest strong conserved interval $J'=(a..c)$ which contains
$I$. Then $J' \subsetneq J$ since otherwise $J$ and $J'$ would
overlap. W.l.o.g. assume that $f_j\neq c$. Then $(f_i..f_k)$
overlaps $J'$, and this contradicts the assumption that $J'$ is strong.
\end{preuve}

We only need three more results before dealing with $b$-nested conserved
intervals.

\begin{lemma}\label{memecontainer}
Let $I$ and $J$ be two conserved intervals. If there exists a frontier $f\in F_I \cap F_J$, 
then $\mbox{Container}(I)$ $=$ $\mbox{Container}(J).$
\end{lemma}

\begin{preuve}
Since both $\mbox{Container}(I)$ and $\mbox{Container}(J)$ are strong,
and since they share a common element, one contains the other. Now, suppose
w.l.o.g. that $\mbox{Container}(I)\subseteq \mbox{Container}(J)$.  By contradiction,
assume that $\mbox{Container}(I)\subsetneq \mbox{Container}(J)$ and let
$\mbox{Container(J)}=(a..c)$. Since $f\in F_J$ and that by Theorem
\ref{theostruc2} we have $F_J=F_{(a..c)}\cap J$, we deduce that $f \in F_{(a..c)}$ and thus
$(a..f)$ and $(f..c)$ are conserved intervals. Now, one of them necessarily
overlaps $\mbox{Container}(I)$, since at least one of $a,c$ is not an endpoint of
$\mbox{Container}(I)$. But this is impossible. Therefore $\mbox{Container}(I)$ $=$
$\mbox{Container}(J).$
\end{preuve}

\begin{lemma}
\label{coro1}
Let $I$ be a (weak or strong) conserved interval with frontier set
$F_I=\{f_1, \ldots, f_k\}$ and let $(a..c)\subseteq I$ be a conserved interval. Then
exactly one of the following cases occurs for the interval $(a..c)$:
\begin{enumerate}
\item either there exists $l$ such that $f_l\in (a..c)$, and then 
there exist $i$ and $j$ such that $(a..c)=(f_i..f_j)$. 
\item or $(a..c)$ contains no frontier of $F_I$, and then there exists $i$ such that
$f_i < a \le c < f_{i+1}$.
\end{enumerate}
\end{lemma}

\begin{preuve} Obviously, the two cases cannot hold simultaneously. Moreover,
in case 2., the deduction is obvious.

Let us focus now on the case 1. Let $(a..c)$ contain some frontier $f_l \in F_I$.
Consider now the two intervals $(f_1..f_l)$ and $(f_l..f_k).$ Then
either {\em (i)} $(a..c)=(f_1..f_k)=I$, or {\em (ii)}
$(a..c)=(f_l..f_l)$, or {\em (iii)} $l=1$ or $l=k$, or {(iv)} $(a..c)$ overlaps
$(f_1..f_l)$ or $(f_l..f_k)$. The two first cases are trivial, let us
consider the two last cases.

In case {\em (iii)} assume w.l.o.g. that $l=1$ (the case $l=k$ is
symmetric). Then $a= f_1$ since $(a..c)\subseteq I$.  Since $(a..c)$
contains the elements of $I$ that are smaller or equal to $c$, then
$(c..f_k)$ is a conserved
interval. Thus $\{f_1,c,f_k\}$ is a frontier set of $I$. If $c\not\in
F_I$ then according to Lemma~\ref{unionfrontier} we have that $\{f_1,c,f_k\}\cup
F_I$ is a frontier-set of $I$ contradicting the maximality of
$F_I$. We deduce that there exists $j$ such that $c=f_j$. 
Now we are done, since $(a..c)=(f_1..f_j)$.

In case {\em (iv)}, $(a..c)$ is necessarily weak since it overlaps another
conserved interval. W.l.o.g. we assume that $(f_1..f_l)$ overlaps $(a..c).$ 
Let us first prove that $f_l$ is also a frontier of $(a..c)$.  Indeed, 
assume {\em a contrario} that $f_l\not \in F_{(a..c)}$, and denote 
$F_{(a..c)}=\{f^*_1, \ldots , f^*_p\}$. With $h\in\{1, 2, \ldots, p\}$,
we have either $f^*_h<f_l$ and by Lemma \ref{lemstru2} for $(f_1..f_l)$ and
$(f^*_h..c)$ we deduce that $(f^*_h..f_l)$ is conserved, or $f_l<f^*_h$ and then
Lemma \ref{lemstru2} for $(a..f^*_h)$ and $(f_l..c)$ we deduce that $(f_l..f^*_h)$
is conserved. Then 
$F_{(a..c)} \cup \{f_l\}$ is a set of frontiers of $(a..c)$, which
contradicts the maximality of $F_{(a..c)}$.
Now, by Lemma \ref{memecontainer} for $\mbox{Container}((a..c))$ (whose
frontier set contains $f_l$ by Theorem \ref{theostruc2}) and $\mbox{Container}(I)$ 
we deduce that $\mbox{Container}((a..c))=\mbox{Container}(I)$. Thus using
Theorem \ref{theostruc2}, we conclude that $(a..c)=(f'_i..f'_j)$ for
some $f'_i,f'_j\in F_{{\mbox{Container}(I)}}$. But since $(a..c)\subseteq I$
and $F_I=F_{{\mbox{Container}(I)}}\cap I$, we are done.
\end{preuve}

The following theorem ensures that in the inclusion tree $T$ of the strong conserved intervals 
of $\PK$, weak intervals are exactly the intervals extending between two frontiers 
of a strong interval. Moreover, each weak interval is uniquely represented in
such a way. In addition, the computation of the tree and of all the frontier sets
is linear.

\begin{theorem}
Let $T$ be the inclusion tree $T$ of strong conserved intervals of a set $\PK$ of permutations.
Then:
\begin{enumerate}
\item  a conserved interval $I$ of $\PK$ is weak if and only if there exists
a strong interval $J$ of $\PK$ and two frontiers $f_i, f_j\in F_J$ such 
that $I=(f_i..f_j)$. Moreover, in this case $J$ is unique.
\item for each strong conserved interval $I$ of $T$ with parent $J$ in $T$, there is a unique conserved
interval $L(I)=(f_i.. f_{i+1})$ defined by  
successive frontiers  in $F_J$ such that $I\subsetneq (f_i..f_{i+1})$. 
\item the tree $T$, the maximal set of frontiers $F_J$ of each strong conserved interval $J$
and the interval $L(I)$ of each strong conserved interval $I\neq (1..n)$ may be computed 
in global $O(n)$ time and $O(n)$ space.
\end{enumerate}
\label{th:arbreT}
\end{theorem}

\begin{preuve}
Concerning affirmation 1, the ''$\Rightarrow$'' part is deduced directly from Theorem~\ref{theostruc2},
whereas the ''$\Leftarrow$'' part is ensured by the definition of a set of frontiers.
Again by Theorem \ref{theostruc2}, we deduce the uniqueness of $J$.

Affirmation 2 results from Lemma \ref{coro1}. According to affirmation 1 in this lemma, $I$ cannot contain
a frontier of $J$, since otherwise $I$ would be of the form $(f_i..f_j)$, with $f_i,f_j\in F_J$, 
and thus would not be strong. Thus, by affirmation 2 in Lemma \ref{coro1}, we deduce the existence of $L(I)$,
which is necessarily unique by the definition of the frontiers.

We focus now on affirmation 3. In \citet{BS06}, a conserved interval is called {\it irreducible} if it cannot 
be written as the union
of smaller conserved intervals. It is easy to notice that the set of irreducible intervals of
size at least two is exactly composed of the intervals $(f_i..f_{i+1})$, where 
$f_i,f_{i+1}$ are two consecutive frontiers of a strong conserved interval of $\PK$.
Indeed, affirmation 1 shows that the only irreducible weak conserved intervals $I=(f_i..f_j)$ 
are those for which $j=i+1$, and obviously the only irreducible strong intervals $I$ are those
with $|F_I|=2$, which are of the form $(f_1..f_2)$, with $F_I=\{f_1,f_2\}$. 

To show affirmation 3, we notice that the number of irreducible intervals is in $O(n)$ [\citet{BS06}],
and that they may be computed in $O(n)$ time and space for an arbitrary number $K$ of permutations using 
generators from \citet{IRconserved}.  Knowing irreducible conserved intervals, the computation of
strong intervals, of their set of frontiers, as well as that of the tree is quite easy. First, one
must plot on the identity permutation the $O(n)$ irreducible intervals of size at least two, by marking the left and
right endpoint of each such irreducible interval. Notice that each element $p$ of the permutation $Id_n$
has at most two marks, the equality occurring only when $p$ is an internal (that is, different from
an endpoint) frontier of a strong interval. We assume the right mark of $p$ (when it exists) 
always precedes the element $p$ whereas the left mark (when it exists) always follows the element $p$, 
so that a left-to-right traversal of $Id_n$ allows to close the interval with right endpoint
$p$ before opening the interval with left endpoint $p$. 

Replacing left and right marks with respectively (square) left and right brackets indexed by their corresponding 
element $p$ on $Id_n$, we obtain an expression $E$ which has correctly nested brackets, 
since irreducible intervals may only overlap on one element. Moreover,  if $I$ and $J$ are strong 
intervals such that $J$ is the parent of $I$, then $I\subsetneq L(I)\subseteq J$ and thus these
intervals are closed exactly in this order during a left-to-right traversal of the expression $E$.
The expression $E$ then allows, during a left-to-right traversal, to discover the strong intervals 
according to a post-order traversal of $T$, which is built on the fly. A strong conserved interval $J$
is obtained by chaining as long as possible neighboring irreducible intervals, i.e. such that the
right bracket of an interval is followed by the endpoint $p$  of the interval and by the left bracket 
of the next interval. Its frontiers  are given by the endpoints of the chained irreducible intervals. 
Also, since $I\subsetneq L(I)$ for all strong intervals $I$, it is easy to identify $L(I)$ since it is 
the interval which closes immediately after $I$ during the traversal. 
\end{preuve}

{\bf Example.} Let $\PK=\{Id_9,P_2\}$, where $P_2=(1,-3,-2,4,5,-8,-7,-6,9)$. The strong intervals
are $(1..9)$ (with frontier set $\{1, 4, 5, 9\}$), $(2..3)$ (with frontier set $\{2,3\}$) and
$(6..8)$ (with frontier set $\{6,7,8\}$). To build $T$, we first plot the irreducible intervals, i.e.
$(1..4), (2..3), (4..5), (5..9), (6..7)$ and  $(7..8)$ on $P_1$, and we obtain the expression:
$$E=1\,\,[_1\,\, -3\,\, [_3\,\,\, ]_2\,\, -2\,\, ]_4\,\ 4\,\, [_4 \,\,\,]_5\,\, 5\,\, [_5\,\, -8\,\, [_8 \,\,\,]_7\,\, -7\,\, [_7 \,\,\,]_6\,\, -6\,\, ]_9\,\, 9\,.$$

\noindent A left-to-right traversal of $E$ allows to find first the interval $(2..3)$ which is included in $(1..4)$,
and thus the node $(2..3)$ of $T$ is built, and a node $x$ starting with $(1..4)$ is created and defined as the parent
of $(2..3)$. Next, the interval $(1..4)$ in $x$ is continued with $(4..5)$ (just change the interval inside
the already existing node), and continued with another interval $(5..t)$, where $t$ is not yet known.
Still,  $(7..8)$ is discovered as a subinterval of $(5..t)$, and it may be continued with $(6..7)$,
thus creating together the interval $(6..8)$ which is another child of $x$. Once this is done, we read
$9$ which indicates that $t=9$. Thus $T$ has three nodes, the root $(1..9)$ (with frontier set $\{1, 4, 5, 9\}$
discovered during the traversal), and its two children $(2..3)$ (with frontier set $\{2,3\}$) and 
$(6..8)$ (with frontier set $\{6,7,8\}$ discovered during the traversal).

\subsection{Properties of $b$-nested conserved intervals}
\label{sect:conservedproperties}

Recall that by definition all $b$-small conserved
intervals are $b$-nested conserved intervals. The
characterization below of $b$-nested conserved intervals has some
similarities with that of $b$-nested common intervals represented in
the $PQ$-tree as $Q$-intervals (see affirmation $(b)$ in Theorem
\ref{th:all}).

 Let $J$ be a conserved interval and let $F_J=\{f_1, f_2, \ldots,
 f_k\}$ be its maximal set of frontiers. We say that $J$ {\em contains a
 $b$-\emph{gap} at position $l$} if $(f_l..f_{l+1})$ is $(b+1)$-large.
 Furthermore, we say that a conserved interval $(a..c)$ {\em falls in the
 gap} between $f_l$ and $f_{l+1}$ of $J$ if $f_l<a\leq c<f_{l+1}$ and
 $\mbox{Container}((a..c))$ has the parent $\mbox{Container}(J)$. In other
 words, $J$ is represented in $T$ by the node $\mbox{Container}(J)$
 and $(a..c)$ is  represented by the node $\mbox{Container}((a..c))$,
 in such a way that the former node is the parent of the latter one.

A $b$-gap at position $l$ is said \emph{good} if it contains at least one
$b$-nested strong conserved interval $I$ with $|I|\geq |(f_l..f_{l+1})|-b$,
or equivalently $|I|\geq f_{l+1}-f_l+1-b$. Then a good $b$-gap is
a $b$-nested conserved interval.

\begin{theorem}\label{thgap}
A conserved interval $J$ of $\PK$ is $b$-nested if and only if it contains no
$b$-gap, or if it contains exactly one good $b$-gap $f_l..f_{l+1}$.
\end{theorem}

\begin{preuve}

We first prove the ''$\Leftarrow$'' part.  Let $J$ be a conserved
interval with maximal set of frontiers $F_J=\{f_1, f_2, \ldots,
f_k\}$. If $J$ contains exactly one $b$-gap, let it be at position
$l$. Otherwise, let us fix arbitrarily $l=1$.  In both cases, $(f_l..f_{l+1})$ is a
conserved $b$-nested interval: either because it is $b$-small, or because
it is a good $b$-gap. Now, for all $j>l+1$ in increasing order, we deduce by
induction that  $(f_l..f_j)$ is a conserved interval (by definition of the set of
  frontiers) containing $(f_l..f_{j-1})$, and thus it is
  $b$-nested. Thus $(f_l..f_k)$ is a conserved $b$-nested interval.
  Similarly, we deduce by induction that $(f_u..f_k)$ is a conserved
  $b$-nested interval, for all decreasing values of $u=l-1, l-2,
  \ldots, 1$. Thus $J=(f_1..f_k)$ is $b$-nested.

Now let us prove the ''$\Rightarrow$'' part, by proving (a) that an
conserved interval with two $b$-gaps or more is not $b$-nested, and (b) that if
a $b$-nested conserved interval contains one $b$-gap then this $b$-gap is good.

{\sf Proof of (a). } Assume by contradiction that $J$ with $F_{J}=\{f_1, f_2, \ldots, f_k\}$ is
conserved $b$-nested and contains two $b$-gaps at initial positions $l$ and $r$
($l < r$). The $b$-nestedness of $J$ implies the existence of a $b$-nested
conserved interval $J'\subsetneq J$ such that $|J'|\geq |J|-b$. Now,
$|J'\cap (f_l..f_{l+1})|\geq 2$, otherwise $J'$ misses at least $b+1$
elements from $(f_l..f_{l+1})$ (which is $(b+1)$-large) and thus 
$|J'|\leq |J|-(b+1)$, a contradiction. Similarly, $|J'\cap (f_r..f_{r+1})|\geq 2$.
We deduce that $J'$ contains both $f_{l+1}$ and $f_{r}$, as well as at least one
additionnal element on the left of $f_{l+1}$ and one additionnal element on the right of 
$f_{r}$. By Lemma \ref{coro1} with $I=J$ and $(a..c)=J'$ we have that $J'=(f_u..f_v)$
with $u\leq l$ and $r+1\leq v$. Thus the existence of the $b$-nested interval
$J$ containing the two $b$-gaps implies the existence of a smaller $b$-nested
interval $J'$ still containing the two $b$-gaps.

Now assume a maximal size series $J_0=J, J_1=J', \ldots$ of $b$-nested conserved 
intervals has been built similarly, each interval being strictly included in the previous one, all containing
the two $b$-gaps. Such a series ends with $(f_l..f_{r+1})$, since
otherwise (if the last interval is larger)  it is possible to construct a smaller $b$-nested interval
included in the last interval and containing $(f_l..f_{r+1}).$ Thus
$(f_l..f_{r+1})$ is $b$-nested. But this is not possible, as it cannot 
strictly contain another $b$-nested common interval of size at least 
$(|f_l..f_{r+1})|-b$. As before, such an interval needs to contain the two gaps, and
it is therefore not strictly included in $(f_l..f_{r+1}).$
 
{\sf Proof of (b).}  Assume now that $J$ with $F_{J}=\{f_1, f_2, \ldots, f_k\}$ is
conserved $b$-nested and has a unique $b$-gap situated at position $l$. As before,
$J$ must contain a $b$-nested conserved interval $J'$ with $|J'|>|J|-b$ and then
$|J'\cap (f_l..f_{l+1})|\geq 2$ implying by Lemma \ref{coro1} that $J'$ must contain
$(f_l..f_{l+1})$. The smallest interval obtained following the same reasoning is then
$(f_l..f_{l+1})$ itself, which must be $b$-nested. As this interval has no internal frontiers
(otherwise $F_{J}$ would not be maximal), any $b$-nested common interval $I=(a..c)$ of size at
least $f_{l+1}-f_l+1-b$ it contains satisfies $f_l<a<c<f_{l+1}$. If $I$ is strong 
then we are done, otherwise $\mbox{Container}(I)$ is strong and has all the required
properties. 
\end{preuve}

\subsection{Computing and counting all $b$-nested conserved intervals}
\label{sect:computingconserved}

Theorem \ref{th:arbreT} allows to count and to enumerate efficiently all
$b$-nested conserved intervals. The computation may be performed, as was
the case for common intervals, in a single post-order traversal of the inclusion tree $T$, 
focusing on each strong conserved interval $I$ of $\PK$. First compute the $b$-gaps of $I$. 
Mark $I$ as $b$-nested if it contains no $b$-gap or one $b$-gap $(f_l..f_{l+1})$, with $f_l,f_{l+1}\in F_I$, 
that is a good one. The latter verification assumes that during the treatment of each child $I'$ of
$I$, if $I'$ is detected as $b$-nested, then $L(I')$ (which is an interval $(f_j..f_{j+1})$ with  
$f_j,f_{j+1}\in F_I$) is marked as good if and only if $L(I')$ is a $b$-gap and $|I'|\geq |L(I')|-b$
(otherwise, $L(I')$ is not marked at all). Now, if $I$ is $b$-nested and $I\neq (1..n)$, 
the same type of mark is performed on $L(I)$ if the conditions are fulfilled.
 
Then, applying Algorithm~\ref{alg2} on each strong conserved interval
$I$ allows to enumerate all the $b$-nested conserved intervals generated by the
frontiers of  $I$, according to Theorem \ref{th:arbreT}. Affirmation 1 in Theorem \ref{th:arbreT}
ensures that each interval is output exactly once.
The running time of Algorithm~\ref{alg2} is clearly linear in the number of intervals output plus
the numbers of children of $I$, yielding a global $O(n+nocc)$ time.

\begin{algorithm}[h!]
\dontprintsemicolon
\caption{Conserved $b$-nested intervals \label{alg2}}
\KwIn{A strong conserved interval $I$, its frontier set $F_I$, the children of $I$ in $T$ marked as nested or not}
\KwOut{The conserved $b$-nested intervals generated by the frontiers of $I$ and strictly included in $I$}
\For{$i$ from $1$ to $|F_I|$}{
$j\gets i+1$\;
goodgaps $\gets$ 0 \hfill{// counts the good $b$-gaps between $f_i$ and $f_j$}\; 
stop $\gets false$\;
\Repeat{{\rm stop} and ($j=|F_I|+1$ or ($i=1$ and $j=|F_i|$))}{
  \If{$f_j - f_{j-1} > b+1$}{
 // Found a  $b$-gap at $j$\;
    \eIf{$(f_j,f_{j+1})$ is a good gap and {\rm goodgaps} $=0$}
        {goodgaps $\gets$ 1\;}
        {stop $\gets true$\;}
   }
   \lIf{not {\rm stop}}{output $(f_i..f_j)$ {\bf end if}}  \hfill{// notice $(f_1..f_{|F_I|})$ is not output}\;
   $j\gets j+1$\;
}
}

\end{algorithm}

To simply count the number of $b$-nested conserved intervals in $I$, we must follow the same
approach as for common intervals. Good and not good $b$-gaps are identified during a search
among the intervals $(f_l..f_{l+1})$, where $F_I=\{f_1, f_2, \ldots,$ $f_k\}$. Then, for
each good $b$-gap we compute the number $l$ (respectively $r$) of successive
$(b+1)$-small intervals $(f_i..f_{i+1})$ on its left (respectively right). We count
the number of $b$-nested conserved intervals containing the good $b$-gap as  
$$l*(r+1)+r.$$ Next, for each maximal set of successive $(b+1)$-small intervals 
$(f_i..f_{i+1})$ we add
$$h*(h-1)/2$$
$b$-nested common intervals, where $h$ is the number of $(b+1)$-small intervals in the set.

All these operations  obvisously take $O(n)$ time and space.

\section{Conclusion}\label{sect:conclusion}
In this paper we introduced the family of $b$-nested common intervals
of $K$ permutations, and showed that it may be computed in time
proportional to its cardinality. This approach extends to any
closed family of intervals that is represented by a $PQ$-tree. We also
show that our approach can be applied to conserved intervals, whose
structure and properties are close but still different from those of common
intervals. The interest of our generalization of common/conserved intervals 
for finding conserved clusters of genes should be attested by further experiments.
Also, other applications may be devised, such as helping the identification of
orthologs/paralogs or defining distances between genomes in an
evolutionary approach. These are the close perspectives of our work.

\bibliographystyle{plainnat}

\begin{thebibliography}{18}
\providecommand{\natexlab}[1]{#1}
\providecommand{\url}[1]{\texttt{#1}}
\expandafter\ifx\csname urlstyle\endcsname\relax
  \providecommand{\doi}[1]{doi: #1}\else
  \providecommand{\doi}{doi: \begingroup \urlstyle{rm}\Url}\fi

\bibitem[Amir et~al.(2007)Amir, Gasieniec, and Shalom]{AmirGS07}
A.~Amir, L.~Gasieniec, and B.~Riva Shalom.
\newblock Improved approximate common interval.
\newblock \emph{Information Processing Letters}, 103\penalty0 (4):\penalty0
  142--149, 2007.

\bibitem[B{\'e}al et~al.(2004)B{\'e}al, Bergeron, Corteel, and
  Raffinot]{BealBCR04}
M.-P. B{\'e}al, A.~Bergeron, S.~Corteel, and M.~Raffinot.
\newblock An algorithmic view of gene teams.
\newblock \emph{Theoretical Computer Science}, 320\penalty0 (2-3):\penalty0
  395--418, 2004.

\bibitem[Bergeron and Stoye(2006)]{BS06}
A.~Bergeron and J.~Stoye.
\newblock On the similarity of sets of permutations and its applications to
  genome comparison.
\newblock \emph{Journal of Computational Biology}, 13\penalty0 (7):\penalty0
  1340--1354, 2006.

\bibitem[Bergeron et~al.(2004)Bergeron, Blanchette, Chateau, and
  Chauve]{BergeronBCC04}
A.~Bergeron, M.~Blanchette, A.~Chateau, and C.~Chauve.
\newblock Reconstructing ancestral gene orders using conserved intervals.
\newblock In \emph{Proceedings of WABI}, volume 3240 of \emph{Lecture Notes in
  Computer Science}, pages 14--25, 2004.

\bibitem[Bergeron et~al.(2008)Bergeron, Chauve, de~Montgolfier, and
  Raffinot]{BCMR05}
A.~Bergeron, C.~Chauve, F.~de~Montgolfier, and M.~Raffinot.
\newblock Computing common intervals of {K} permutations, with applications to
  modular decomposition of graphs.
\newblock \emph{SIAM Journal of Discrete Mathematics}, 22\penalty0
  (3):\penalty0 1022--1039, 2008.

\bibitem[Blin et~al.(2010)Blin, Faye, and Stoye]{BlinFS10}
G.~Blin, D.~Faye, and J.~Stoye.
\newblock Finding nested common intervals efficiently.
\newblock \emph{Journal of Computational Biology}, 17\penalty0 (9):\penalty0
  1183--1194, 2010.

\bibitem[Fertin and Rusu(2011)]{FR}
G.~Fertin and I.~Rusu.
\newblock Computing genomic distances: An algorithmic viewpoint.
\newblock In \emph{Algorithms in Computational Molecular Biology: Techniques,
  Approaches and Applications, {\rm M. Elloumi, A. Y. Zomaya} {\it eds.}},
  pages 773--797. Wiley Series in Bioinformatics, 2011.

\bibitem[Galperin and Koonin(2000)]{galperin2000s}
M.~Y Galperin and E.~V Koonin.
\newblock Who's your neighbor? new computational approaches for functional
  genomics.
\newblock \emph{Nature Biotechnology}, 18\penalty0 (6):\penalty0 609--613,
  2000.

\bibitem[Hoberman and Durand(2005)]{HD05}
R.~Hoberman and D.~Durand.
\newblock The incompatible desiderata of gene cluster properties.
\newblock In \emph{Proceedings of RECOMB-CG}, volume 3678 of \emph{Lecture
  Notes in Computer Science}, pages 73--87, 2005.

\bibitem[Kurzik-Dumke and Zengerle(1996)]{The10}
U.~Kurzik-Dumke and A.~Zengerle.
\newblock Identification of a novel drosophila melanogaster gene, angel, a
  member of a nested gene cluster at locus 59f4,5.
\newblock \emph{Biochim Biophys Acta}, 1308\penalty0 (3):\penalty0 177--81,
  1996.

\bibitem[Landau et~al.(2005)Landau, Parida, and Weimann]{landau2005using}
G.~M. Landau, L.~Parida, and O.~Weimann.
\newblock Using {$PQ$}-trees for comparative genomics.
\newblock In \emph{Proceedings of CPM}, volume 3537 of \emph{Lecture Notes in
  Computer Science}, pages 128--143, 2005.

\bibitem[Lathe et~al.(2000)Lathe, Snel, and Bork]{lathe2000gene}
W.~C Lathe, B.~Snel, and P.~Bork.
\newblock Gene context conservation of a higher order than operons.
\newblock \emph{Trends in Biochemical Sciences}, 25\penalty0 (10):\penalty0
  474--479, 2000.

\bibitem[Parida(2006)]{Parida06}
L.~Parida.
\newblock Gapped permutation patterns for comparative genomics.
\newblock In \emph{Proceedings of WABI}, volume 4175 of \emph{Lecture Notes in
  Computer Science}, pages 376--387, 2006.

\bibitem[Pasek et~al.(2005)Pasek, Bergeron, Risler, Louis, Ollivier, and
  Raffinot]{Pasek2005}
S.~Pasek, A.~Bergeron, J.~L. Risler, A.~Louis, E.~Ollivier, and M.~Raffinot.
\newblock {I}dentification of genomic features using microsyntenies of domains:
  domain teams.
\newblock \emph{Genome Research}, 15\penalty0 (6):\penalty0 867--874, June
  2005.

\bibitem[Rusu(2012)]{IRconserved}
I.~Rusu.
\newblock New applications of interval generators to genome comparison.
\newblock \emph{Journal of Discrete Algorithms}, 10:\penalty0 123--139, 2012.

\bibitem[Rusu(2013)]{IR}
I.~Rusu.
\newblock Min{M}ax-{P}rofiles: A unifying view of common intervals, nested
  common intervals and conserved intervals of {$K$} permutations.
\newblock arXiv:1304.5140, submitted, 2013.

\bibitem[Tamames(2001)]{tamames2001evolution}
J.~Tamames.
\newblock Evolution of gene order conservation in prokaryotes.
\newblock \emph{Genome Biology}, 2\penalty0 (6):\penalty0 R0020, 2001.

\bibitem[Uno and Yagiura(2000)]{UnoYagura}
T.~Uno and M.~Yagiura.
\newblock Fast algorithms to enumerate all common intervals of two
  permutations.
\newblock \emph{Algorithmica}, 26:\penalty0 290--309, 2000.

\end{thebibliography}

\end{document}